\documentclass{ifacconf}

\usepackage{amsmath} 
\usepackage{amssymb}  
\usepackage{MnSymbol}
\usepackage{subfig}
\usepackage{xcolor}

\usepackage{graphicx}      
\usepackage{natbib}        

\usepackage{tikz}
\usepackage{pgfplots}
\usepackage{verbatim}
\allowdisplaybreaks
\pgfplotsset{every axis y label/.style={at={(ticklabel cs:0.5)},rotate=90,anchor=center}}
\newtheorem{corollary}[thm]{Corollary}

\begin{document}
\begin{frontmatter}

\title{Geometric tracking control for a nonholonomic system: a spherical robot} 


\author[First]{Sneha Gajbhiye} 
\author[First]{Ravi N. Banavar} 

\address[First]{Systems and Control Engineering, Indian Institute of Technology Bombay,
India, 400076. (e-mail: sneha@sc.iitb.ac.in, banavar@iitb.ac.in)}

\begin{abstract}  
This paper presents tracking control laws for two different objectives of a nonholonomic system - a spherical robot - using a geometric approach. The first control law addresses orientation tracking using a modified trace potential function. The second law addresses contact position tracking using a $right$ transport map for the angular velocity error. A special case of this is position and reduced orientation stabilization. Both control laws are coordinate free. The performance of the feedback control laws are demonstrated through simulations.

\end{abstract}


\end{frontmatter}

\section{Introduction}
The problem of tracking of nonholonomic systems is a challenging one in control theory. Applications include robotics, rolling and locomotive mechanisms. A better understanding of  the system's intrinsic properties simplify the control synthesis. Geometric control theory plays an important role in accomplishing such design strategies, see \citep{book_isidori}, \citep{Zenkov}, \citep{ostrowski_thesis}. In this paper we study the tracking problem of one such nonholonomic system - a \textit{spherical robot}.

A spherical mobile robot is a spherical shell actuated by a driving mechanism mounted inside to make the shell roll. In this paper we consider the driving actuators as three rotors. So the robot has three input degrees of freedom (rotors) which are used to control two translation and three rotational degrees of freedom (shell). Several modeling approaches and motion planning algorithms have been proposed for the spherical robot to achieve desired orientation and position,
 see \citep{joshi_banavar}, \citep{reg_n_chaotic_2013}, \citep{mukherjee1999}, \citep{bicchi1997}, \citep{Zhan_2008}, that are solely based on coordinate dependent approach like quaternions and Euler parametrizations. Geometric control addressed the development of control laws for systems evolving on manifolds in a coordinate free setting. Recently, \cite{schneider} has derived the dynamic model of the  Chaplygin's sphere using geometric mechanics and presented  orientation stabilization of a Chaplygin's sphere with a rotor by the controlled Lagrangian matching condition. In \citep{karimpur_2012}, \citep{muralidharan_mahindrakar}, the authors address the control methods based on quaternions and stereographic projection respectively. In \cite{karimpur_2012}, the authors applied backstepping to achieve position stabilization and tracking by expressing attitude in quaternion representation. In \citep{shen2008}, the authors propose motion planning algorithms using symmetric products on manifold (Lie group) to achieve position convergence with arbitrary orientation and vice versa. As spherical robot is a nonholonomic system, it fails to satisfy a necessary condition for asymptotic stabilization on $SO(3) \times \mathbb{R}^{2}$ by a continuous feedback law, see Brockett \citep{brockett}. Due to this negative result, point-to-point stabilization of position and orientation of spherical robot through continuous state-feedback is not possible. In \citep{muralidharan_mahindrakar}, the authors consider stereographic projection map and design smooth kinematic control law to achieve position tracking and position with reduced attitude stabilization. 

The contribution of this paper is to present two geometric control laws to achieve two different objectives: 1) tracking of a desired orientation trajectory; 2) contact position trajectory tracking asymptotically. The intermediate result of contact position tracking law is position and reduced attitude stabilization. The use of the transport map for velocity error on $SO(3)$ gives a better and complete understanding of the nonholonomic constraint in case of position tracking. For orientation tracking, a potential function which is the trace of the relative orientation and desired orientation, is constructed. The stability result is derived using Lyapunov direct method \citep{nijmeijer} which is recently restated in \citep{bullo_stability} to achieve asymptotic stability. While the notion of transport map in velocity error has been considered (see, \citep{book_bullo} for tracking of fully actuated and \citep{taeyoung_lee} for underactuated systems), this is the first instance where such a treatment is considered in the presence of a nonholonomic constraint and with underactuation. 

The model is derived using Lagrangian reduction defined on a symmetry group. The well developed theory on geometric nonholonomic mechanics is presented in \citep{marsden}, \citep{CMR}, \citep{CHMR}, \citep{BKMM}, \citep{bloch2003}. By symmetry we can study the dynamics of a mechanical system on a reduced space and the reduced equations are in the Euler-Poincar\'{e} form. Due to nonholonomic constraints the system may or may not have full symmetry as in the case of the rigid body with gravitational field, for example, a heavy top; and the Euler-Poincar\'{e} equation will depend on an advection term \citep{schneider}. In this paper we follow this modelling tool and derive the reduced equations of motion. The paper is organized as follows: In Section 2 we present the description and modelling of the spherical robot using Lagrangian reduction theory. In section 3 we formulate the control problem for orientation tracking and then position tracking and axis stabilization. We identify this stabilization as position and axis stabilization. Section 4 follows with the concluding remarks on the above control strategies.

\section{Description of a spherical robot}
\begin{figure}[h]
\centering
\includegraphics[scale=0.3]{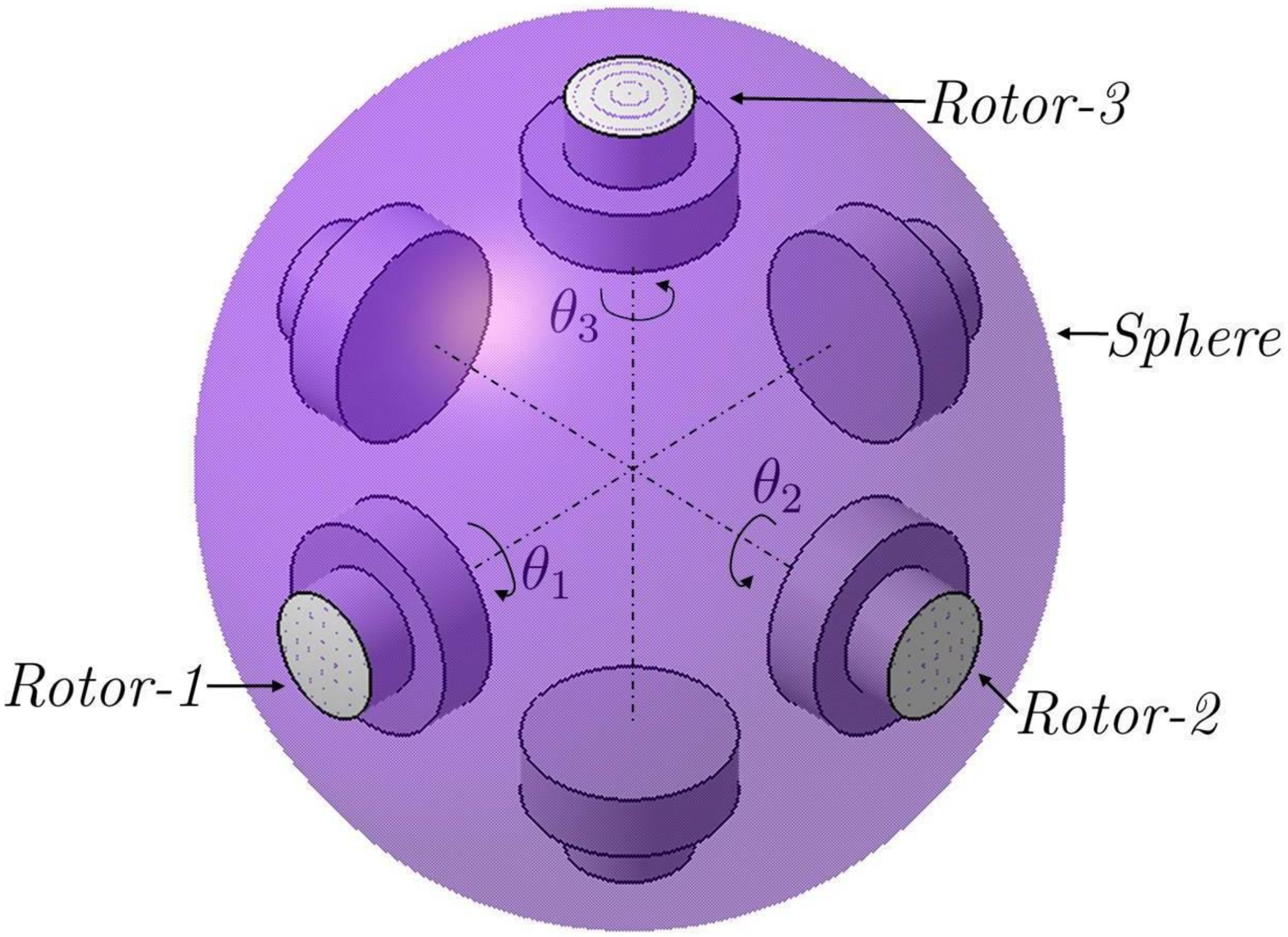}
\caption{Spherical robot on horizontal plane}
\label{rotor_fig}
\end{figure}

Consider a spherical mobile robot  with internal rotors which can roll without slipping on a flat surface under a uniform gravitational field. All the three rotors are placed along three mutually orthogonal axes of the sphere-body frame, as shown in Fig. (\ref{rotor_fig}). To balance the mass symmetrically, the rotor is placed on one side and a dead weight is placed on the diametrically opposite side. All the rotors and dead weights are placed such that the center of mass of the robot coincides with the geometric center of the sphere. 
Let the sphere body coordinate frame be located with its origin at the center of the sphere. Let $\mathbf{x} \in R^{3}$ be the position of the center of the sphere in an inertial frame, and let $R_{s}\in SO(3)$ be the rotation matrix which maps from the sphere body coordinate frame to the inertial coordinate frame. The relative motion of three rotors with respect to the sphere body frame is given by generalized shape coordinates $\theta_{i} \in S^{1}$,  where $i = 1,2,3$. Hence, the configuration space is $Q={\mathbb R}^{2} \times SO(3)\times Q_{s}$, where $Q_{s} = S^{1} \times S^{1} \times S^{1}$. The following notation is adopted here:
\begin{itemize}
\item $(\hat{e}_{1},\hat{e}_{2},\hat{e}_{3})$ - Unit vectors in inertial frame,
\item $\omega_{s}^{I}$, $\omega_{s}^{s}$ - Angular velocity of the sphere in inertial frame and sphere frame respectively; $\dot{\theta}_{i}$ - Angular velocity of the $i^{th}$ rotor,
%
%
\item $m_{s}, m_{i}$ - Mass of the sphere and $i^{th}$ rotors; $\mathbb{I}^{s} = diag(I_{s},I_{s},I_{s})$ - Inertia matrix of the sphere without rotors about its center of mass in sphere frame and $J^{i} $ - Moment of inertia of the rotors about the three principal axes, i.e. $J^{1}=diag(J_{a},J_{b},J_{b});J^{2}=diag(J_{b},J_{a},J_{b});J^{3}=diag(J_{b},J_{b},J_{a})$ with $J_{a}=2J_{b}$.
\end{itemize}
The Lagrangian of the system consists only of kinetic energy and is given as
\begin{align*}
L = \frac{1}{2} \left( m_{T} \|\dot{\mathbf{x}}\|^{2} +  \sum_{i=1}^{3} (I_{i} \omega_{i}^{2} + J_{a}(\omega_{i} + \dot{\theta}_{i})^{2}) \right)
\end{align*}
where $\omega_{s}^{s} \triangleq [\omega_{1},\omega_{2},\omega_{3}]$, $m_{T}= (m_{s} + \sum_{i = 1}^{3} m_{i})$ and $I_{i} = I_{s} + 2 J_{b}$. The Lagrangian is now expressed as 
\begin{equation}\label{lagrange}
L = \frac{1}{2} m_{T} \|\dot{\mathbf{x}}\|^{2} + \frac{1}{2} \omega_{s}^{T}I^{s} \omega_{s}^{s} + \frac{1}{2}\left( \dot{\Theta} + \omega_{s}^{s}\right)^{T}\mathbb{J}\left(\dot{\Theta} + \omega_{s}^{s}\right)
\end{equation}
where $\mathbb{J} = diag(J_{a},J_{a},J_{a})$, $I^{s}=dia(I_{1},I_{2},I_{3})$, $\&$ $\dot{\Theta}=(\dot{\theta}_{1},\dot{\theta}_{2},\dot{\theta}_{3})$. The rolling without slipping assumption on the robot yields a nonholonomic constraint given as
\begin{equation}\label{constr1}
\dot{\mathbf{x}}- \omega_{s}^{I}\times r\hat{e}_{3} =0.
\end{equation}
\subsection{Dynamics of the spherical robot} 
The configuration space $Q$ is a smooth manifold, $TQ$ is the velocity space called the tangent bundle and a smooth distribution $\mathcal{D} \subset TQ$ defines the constraints, the set of admissible velocities. 
With the Lagrangian $L$ defined in (\ref{lagrange}) and distribution $\mathcal{D}\subset TQ$ satisfying the constraint (\ref{constr1}), let $G =SO(3) \times \mathbb{R}^{3}$ be a group with its Lie algebra $\mathfrak{g} = \mathfrak{so}(3) \times \mathbb{R}^{3}$, where $\mathfrak{so}(3)$ is the Lie algebra of $SO(3)$. The group action of $G$ on $Q$ is given by $\Phi: G \times Q \mapsto Q$; $\Phi_{(\bar{R}_{s},\bar{b})}(R_{s},b,\Theta) = (\bar{R}_{s}R_{s},\bar{R}_{s}b + \bar{b}, \Theta)$. It is seen that the Lagrangian $L$ and distribution $\mathcal{D}$ is invariant with respect to the subgroup $G_{\hat{e}_{3}}$ of $G$ given as
$ G_{\hat{e}_{3}}=\lbrace (R_{s},b)\in G \vert R_{s}^{T} \hat{e}_{3}=\hat{e}_{3} \rbrace = SO(2)\times \mathbb{R}^{2}$. When the Lagrangian $L$ and the distribution $\mathcal{D}$ are invariant under the action of the subgroup group $G_{\hat{e}_{3}}$, the system is reduced to the space $TQ/G_{\hat{e}_{3}}$ and the Lagrangian is termed as the reduced Lagrangian $l$. Define 
\begin{equation}\label{advection}
\bar{Y}= R_{s}^{T} \dot{\mathbf{x}} \quad \quad \Gamma\triangleq R_{s}^{T}\hat{e}_{3},
\end{equation}
where $\bar{Y}$ is the velocity of the contact point in the sphere frame and $\Gamma$ is called an advected variable \citep{lhmnlc2012}. The reduced Lagrangian $l: TQ/G_{\hat{e}_{3}} \longrightarrow \mathbb{R} $ 
\begin{equation}\label{equ2}
l = \frac{1}{2} m_{T} \|\bar{Y}\|^{2} + \frac{1}{2} \omega_{s}^{s}\cdot (I^{s} + \mathbb{J}) \omega_{s}^{s} + \frac{1}{2}\left( \dot{\Theta} \cdot \mathbb{J}\dot{\Theta} + 2 \omega_{s}^{s} \cdot \mathbb{J}\dot{\Theta}\right) 
\end{equation}
and the rolling constraint is now expressed in the sphere body coordinate frame as $\bar{Y} = r \widehat{\omega}_{s}^{s}\Gamma$, 
%
where $\bar{Y}$, $\Gamma$ $\in \mathbb{R}^{3}$ and $\widehat{\omega}_{s}^{s} = R_{s}^{T}\dot{R}_{s} \in \mathfrak{so}(3)$ is the (left-invariant) sphere-body angular velocity. Substituting $\bar{Y}$ in $l$, the system is reduced to the quotient space $\mathcal{D}/G_{\hat{e}_{3}}$ given by the reduced-constraint Lagrangian $l_{c}$ as
$$l_{c} = + \frac{1}{2} \omega_{s}^{s}\cdot (- m_{T}r^{2}\widehat{\Gamma}\widehat{\Gamma} + I^{s} + \mathbb{J}) \omega_{s}^{s} + \frac{1}{2}\left( \dot{\Theta} \cdot \mathbb{J}\dot{\Theta} + 2 \omega_{s}^{s}\cdot \mathbb{J}\dot{\Theta}\right)$$
Due to subgroup symmetry, there is an advection dynamic and differentiating (\ref{advection}) it is calculated as
\begin{equation}\label{final_adv}
\dot{\Gamma} = - \omega_{s}^{s} \times \Gamma.
\end{equation}
The dynamics is calculated using the intermediate theorem given by \citep{schneider}. The equations of motion is given by the Euler-Poincar\'{e} equation for the group variable $R_{s}$ and the Euler-Lagrange equation for the shape variable $\Theta$.
%
Let $\Pi_{s} = \frac{\partial l_{c}}{\partial \omega_{s}^{s}} $ be the angular momentum of the sphere(momentum conjugate to $\omega_{s}^{s}$) and $\Pi_{i}$ be the angular momentum of the $i^{th}$ rotor,the dynamics is given as, 
\begin{align}\label{equ3}
\dot{\Pi}_{s} = \Pi_{s} \times \omega_{s}^{s}; \quad \quad \dot{\Pi}_{i} = u,
\end{align}
Recasting the dynamic equation \eqref{equ3} as,
\begin{align}\label{reduced_dyanamic}
\begin{split}
& M(\Gamma) \dot{\omega}_{s}^{s} = (I^{s} \omega_{s}^{s} + \mathbb{J}\dot{\Theta}) \times \omega_{s}^{s} - u,\\
& \mathbb{J}\dot{\omega}_{s}^{s} + \mathbb{J}\ddot{\Theta} = u.
\end{split}
\end{align}
where $M(\Gamma)= I^{s} - m_{T}r\widehat{\Gamma}^{T}\widehat{\Gamma}$ and using the solution $\omega_{s}^{s}$ of the equation (\ref{reduced_dyanamic}), we can find the curve $R_{s}(t)$ by solving the reconstruction equation 
\begin{equation}\label{rec}
\dot{R}_{s}(t) = R_{s}(t)\widehat{\omega}_{s}^{s} \text{  with  }~~R_{s}(0)=R_{s_{0}}.
\end{equation}
Hence, equations (\ref{constr1}), (\ref{reduced_dyanamic}) and (\ref{final_adv}), together with the reconstruction equation (\ref{rec}), give the complete dynamics of the spherical robot. If $u^{i}=0$, it is easily seen that any configuration is an equilibrium and hence the equilibrium manifold is the whole configuration manifold $Q$. By expressing the control in terms of the gradient of a potential function (or error function), the equilibrium can be changed to any desired point. A similar procedure is followed in the next two sections to achieve tracking. 
\subsection*{Remarks on controllability}
The controllability for the three rotor case has been analysed by \cite{reg_n_chaotic_2013}, \cite{joshi_banavar} in the literature. One can use fiber configuration controllability definition to check the controllability. This controllability has been studied for the Chaplygin's sphere with rotors in \cite{shen2008}, \cite{karimpur_2012}. We will mention this result and then design stabilization/tracking control laws for our system. To check the controllability, equation (\ref{reduced_dyanamic}) is cast in an affine-control form as
\begin{equation}\nonumber
\dot{q} = f(q) + g(q)u
\end{equation}
where, $q = (\omega_{s}^{s}, \dot{\Theta})$, the drift vector field $f =0$ and the control vector fields $ g = [g_{1}~g_{2}~g_{3}]$, are expressed as 
\begin{equation}
g_{i} = \begin{bmatrix}
- (I^{s} + \mathbb{J} - m_{T}r\widehat{\Gamma}^{T}\widehat{\Gamma})^{-1} \mathbb{J} \Delta^{-1}\\
\Delta^{-1}
\end{bmatrix} \hat{e}_{i}
\end{equation}
where $\Delta = \mathbb{J} - \mathbb{J}(I^{s} + \mathbb{J} - m_{T}r\widehat{\Gamma}^{T}\widehat{\Gamma})^{-1} \mathbb{J}$. If  $v = \Delta~u$ is an equivalent control input, where $v = (v_{1},v_{2},v_{3})$ is a transformed control input; then the control vector fields on $SO(3) \times Q_{s}$ are written as 
\begin{equation}
g_{i} \simeq \begin{bmatrix}
- A_{i}(\Gamma) \\
\hat{e}_{i}
\end{bmatrix}
\end{equation}
where $A_{i}$ is the $i^{th}$ column of $A = (I^{s} - m_{T}r\widehat{\Gamma}^{T}\widehat{\Gamma})^{-1} \mathbb{J}$. Let $\pi_{s} = R_{s} \Pi_{s}$, then from  (\ref{reduced_dyanamic}), which is the Euler-Poincar\'{e} form, we see 
\begin{equation}\label{conserve1}
\frac{d}{dt}\left( R_{s} \Pi_{s} \right)  = 0.
\end{equation}%
that is, the inertial momentum $\pi_{s}$ is conserved. Suppose that the system is initially at equilibrium, then $\Pi_{s} = 0$ and therefore $\omega_{s} = - (I^{s} + \mathbb{J} - m_{T} r \widehat{\Gamma}\widehat{\Gamma})^{-1}\mathbb{J}\dot{\Theta} = - A(\Gamma) \dot{\Theta}$, where $A(\Gamma)$ is called as \textit{mechanical connection} \cite{schneider}. From (\ref{advection}), $\bar{Y} = r A(\Gamma)\dot{\Theta} \times \Gamma$, then the control vector fields for the complete configuration $SO(3) \times Q_{s} \times \mathbb{R}^{2}$ are expressed as 
\begin{equation}
\bar{g}_{i} \simeq \begin{bmatrix}
- A(\Gamma)\hat{e}_{i} \\
\hat{e}_{i} \\
r A(\Gamma)\hat{e}_{i} \times \Gamma
\end{bmatrix}
\end{equation}
The iterative Lie brackets $\{\bar{g}_{1}, \bar{g}_{2}, \bar{g}_{3}, [\bar{g}_{1},\bar{g}_{2}], [\bar{g}_{1},\bar{g}_{3}] \}$ span the tangent space of $SO(3) \times \mathbb{R}^{2}$ (termed as the \textit{fiber configuration}) at any configuration. Hence, the system is \textit{fiber} configuration accessible at any configuration and therefore fiber configuration controllable. 
\section{Orientation trajectory tracking}
The control objective here is to design a feedback control law which tracks a desired orientation trajectory $R_{d}(t)$.  The rotational system dynamics described by (\ref{reduced_dyanamic}) and (\ref{rec}) can be expressed in the standard control form with $q = (R_{s},\omega_{s}^{s})$ as
\begin{equation}\label{control_equation2}
\dot{q} = f(q) + g(q)u
\end{equation}
where
\begin{align*}
f = \begin{bmatrix}
R_{s}\widehat{\omega}_{s}^{s} \\
M^{-1}\left(  (I^{s}\omega_{s}^{s} + \mathbb{J}\dot{\Theta}) \times \omega_{s}^{s}\right) 
\end{bmatrix}; \quad
g = [g_{1}~ g_{2}~ g_{3}] = \begin{bmatrix}
\mathbf{0} \\
-b_{i}
\end{bmatrix}. 
\end{align*}
where for notational simplification we write $M(\Gamma)=M$ and $b_{i}'s$ are the columns of $M^{-1}$.
We now define a scalar valued potential function to achieve this objective and then prove the stability of the system. Subsequently, we add a damping term to get asymptotic convergence to the equilibrium. Let $V: Q \longrightarrow \mathbb{R}$ be an error function about $R_{d}(t)\in SO(3)$ constructed by a modified trace function as
\begin{equation}
V(R_{s}) = \mbox{trace}(K_{p}(I_{3\times 3} -  R_{d}^{T}R_{s})).
\end{equation}
where $K_{p} = diag(\lambda_{1},\lambda_{2},\lambda_{3})$ with $\lambda_{1},\lambda_{2},\lambda_{3}>0$ and $\lambda_{1} \neq \lambda_{2} \neq \lambda_{3}$. The modified trace function was first employ by \citep{chillingworth_marsden_wan} for the purpose of feedback stabilization. Define $\widehat{\omega}_{d}^{s} = R_{d}^{T}\dot{R}_{d}$ and set $R_{e} = R_{d}^{T}R_{s}$ and the error in angular velocity as $e_{\omega} \triangleq \omega_{s}^{s} - R_{e}^{T}\omega_{d}^{s}$. Taking time derivative of $V$,
\begin{align}\nonumber
\begin{split}
& \frac{d V}{dt}  =\mbox{trace}( K_{p}(R_{d}^{T}\dot{R}_{d}R_{d}^{T}R_{s} -  R_{d}^{T} \dot{R}_{s})) \\
& = \mbox{trace}( K_{p} R_{e}(R_{e}^{T} \widehat{\omega}_{d}^{s}R_{e} - \widehat{\omega}_{s}^{s})) = - \mbox{trace}(k_{p}R_{e}\widehat{e}_{\omega})\\
& = - \frac{1}{2} \mbox{trace}([\mbox{skew}(K_{p}R_{d}^{T}R_{s}) + \mbox{sym}(K_{p}R_{d}^{T}R_{s})] \widehat{e}_{\omega})\\
&  = - \frac{1}{2} \mbox{trace}(\mbox{skew}(K_{p}R_{d}^{T}R_{s}) (\widehat{e}_{\omega}))
\end{split}
\end{align}
and from the equality $\mbox{trace}(\widehat{x}\widehat{y}) = - 2 x \cdot y $, where $\widehat{\cdot} : \mathbb{R}^{3} \longrightarrow \mathfrak{so}(3)$ is a \textit{hat} map and $(\cdot)^{\vee} : \mathfrak{so}(3)\longrightarrow \mathbb{R}^{3}$ is a \textit{breve} map (inverse of hat map), it follows that
\begin{align}\nonumber
\begin{split}
 & \dot{V}  =  \mbox{skew}(K_{p}R_{d}^{T}R_{s})^{\vee} \cdot e_{\omega} = (\sum_{i = 1}^{3} \lambda_{i} R_{s}^{T}\hat{e}_{i} \times R_{d}^{T}\hat{e}_{i})\cdot e_{\omega} = dV \cdot e_{\omega},
\end{split}
\end{align} 
where $dV$ can also be termed as differential of $V$ with respect to $R_{s}$.

We compute the feedforward (FF) control term which tracks the desired velocity and add the proportional-derivative (PD) term to stabilize/track the orientation asymptotically. The velocity error has geometric interpretation since $\widehat{\omega}$ is Lie algebraic element. Since $\dot{R}_{s}$ and $\dot{R}_{d}$ are the two velocities taking values in different tangent spaces, to define the error velocity we need to compare tangent vectors in the same tangent space. This can be achieved by the transport map $\tau$ as defined in [\citep{book_bullo}, \S 11]. If $\dot{R}_{s} \in T_{R_{s}}SO(3)$ and $\dot{R}_{d} \in T_{R_{d}}SO(3)$ are the two vectors at the points $R_{s}$ and $R_{d}$ respectively, then a right transport map $\tau(R_{s},R_{d})$ transforms $\dot{R}_{d}$ into a vector at $T_{R_{s}}SO(3)$ and the error is expressed as
\begin{align}\nonumber
\begin{split}
& \dot{R}_{s}  - \tau(R_{s},R_{d}) (\dot{R}_{d}) = \dot{R}_{s} - \dot{R}_{d}(R_{d}^{T}R_{s}),\\
& = R_{s}R_{s}^{T} \dot{R}_{s} - (R_{s}R_{s}^{T})(R_{d}R_{d}^{T})\dot{R}_{d}(R_{d}^{T}R_{s}),\\
& = R_{s}\widehat{\omega}_{s}^{s} - R_{s}(R_{s}^{T}R_{d})\widehat{\omega}_{d}^{s}(R_{d}^{T}R_{s})= R_{s}[\widehat{\omega}_{s}^{s} - (R_{e}^{T}\omega_{d}^{s})^{\wedge}] = R_{s}\widehat{e}_{\omega}
\end{split}
\end{align}
Now, the feedforward control term is calculated by taking the covariant derivative of the transport map along $\omega_{s}^{s}$. Associated with a Riemannian manifold is the notion of the affine connection $\nabla$ that defines the covariant derivative. For details on Riemannain manifolds and affine differential geometry one can refer to \citep{do_carmo}, \citep{book_bullo}. For a given affine connection $\nabla$, two vector fields $X_{\xi},X_{\eta}$ with $\xi, \eta \in \mathfrak{g}$ its Lie algebra, the covariant derivative is defined as $\nabla_{X_{\xi}}X_{\eta}$. If $X_{\xi},X_{\eta}$ are left-invariant vector fields on $Q$, then the covariant derivative is $\nabla_{X_{\xi}}X_{\eta} = \frac{d}{dt}X_{\eta} + \overset{\mathfrak{g}}{\nabla}_{\xi}\eta$, 
%
where $\overset{\mathfrak{g}}{\nabla}_{\xi}\eta$ is a bilinear map defined as 
\begin{equation}\label{covariant}
\overset{\mathfrak{g}}{\nabla}_{\xi}\eta = M^{-1}\left( \frac{1}{2}(\xi \times M\eta) + \frac{1}{2}(\eta \times M\xi) \right). 
\end{equation}
In our case $X_{\xi} = \dot{R}_{s}$, $X_{\eta} = \tau(R_{s},R_{d})\dot{R}_{d}$ and $\mathfrak{g} = \mathfrak{so}(3)$ with $\xi=\widehat{\omega}_{s}^{s}$ and $\eta = (R_{d}^{T}\omega_{d}^{s})^{\wedge}$. With this the $f_{FF}$ is calculated as $f_{FF} = M \left( \nabla_{\dot{R}_{s}}\tau(R_{s},R_{d})\dot{R}_{d} \right), $
\begin{align}
& = M \left( \frac{d R_{e}^{T}}{dt} \omega_{d}^{s} + \overset{\mathfrak{so}(3)}{\nabla}_{\widehat{\omega}_{s}^{s}}(R_{e}^{T}\omega_{d}^{s})^{\wedge} \right), \nonumber \\
& = M \left( \frac{d R_{e}^{T}}{dt}\omega_{d}^{s} + R_{e}^{T} \frac{d \omega_{d}^{s}}{dt} + \overset{\mathfrak{so}(3)}{\nabla}_{\widehat{\omega}_{s}^{s}}(R_{e}^{T}\omega_{d}^{s})^{\wedge} \right), \nonumber \\
& =  M \left( (\omega_{s}^{s} \times R_{e}^{T}\omega_{d}^{s}) + R_{e}^{T}\dot{\omega}_{d}^{s} +\overset{\mathfrak{so}(3)}{\nabla}_{\widehat{\omega}_{s}^{s}}(R_{e}^{T}\omega_{d}^{s})^{\wedge} \right).\nonumber 
\end{align}
From (\ref{covariant}) calculating the bilinear map and therefore 
\begin{align}
f_{FF} &= M \left(\frac{1}{2} M^{-1}(\omega_{s}^{s} \times M R_{e}^{T}\omega_{d}^{s}) - \frac{1}{2} M^{-1}(M\omega_{s}^{s} \times R_{e}^{T}\omega_{d}^{s})\right) \nonumber \\
& \quad + M (\omega_{s}^{s} \times R_{e}^{T}\omega_{d}^{s}) + M R_{e}^{T}\dot{\omega}_{d}^{s}. \label{feedfrwd}
\end{align}
\begin{thm}
Under the feedback torque $u(R_{s}) = dV(R_{s}) - f_{FF}$ the closed loop system (\ref{control_equation2}) is Lyapunov stable about $(R_{d}, \omega_{d}^{s})$.
\end{thm}
\textit{Proof}: Define the function $H : TQ \longrightarrow \mathbb{R}$
\begin{equation}\label{lyapunov_function}
H(R_{s},\omega_{s}^{s}) = V(R_{e}) + \frac{1}{2}\| e_{\omega}\|_{M}^{2} = V(R_{e}) + \frac{1}{2}\mathbb{G}(I)(e_{\omega},e_{\omega}),
\end{equation}
where $\mathbb{G}(I) = M$ is the Riemanian metric on $Q$. Since, $V$ is an error function and $M>0$, it follows that the function $H$ is locally positive definite around $(R_{d}, \omega_{d}^{s})$. It follows
\begin{align}
& \frac{d}{dt}H(\mathbf{x},\omega_{s}^{s}) = \frac{d}{dt}V + \mathbb{G}(I)(e_{\omega},\nabla_{\omega_{s}^{s}}e_{\omega}), \nonumber \\
& = \dot{V} + \mathbb{G}(I)(e_{\omega},\nabla_{\omega_{s}^{s}}\omega_{s}^{s} - \nabla_{\omega_{s}^{s}}R_{e}^{T} \omega_{d}^{s}), \nonumber \\
& = \dot{V} + \langle e_{\omega}, M\left( \frac{d}{dt}\omega_{s}^{s} + \overset{\mathfrak{so}(3)}{\nabla}_{\widehat{\omega}_{s}^{s}}\widehat{\omega}_{s}^{s} \right)  \rangle - \langle e_{\omega}, M \nabla_{\omega_{s}^{s}}R_{e}^{T} \omega_{d}^{s} \rangle, \nonumber  \\
& = \dot{V} + \langle e_{\omega}, M\dot{\omega}_{s}^{s} + ( \omega_{s}^{s} \times (I^{s}\omega_{s}^{s} + \mathbb{J} \dot{\Theta})) - M \nabla_{\omega_{s}^{s}}R_{e}^{T} \omega_{d}^{s} \rangle, \nonumber \\
& = \dot{V} + \langle e_{\omega}, - u \rangle - \langle e_{\omega}, M \nabla_{\omega_{s}^{s}}R_{e}^{T} \omega_{d}^{s} \rangle, \nonumber \\
& = dV \cdot e_{\omega} + \langle e_{\omega}, -dV + f_{FF}  \rangle - \langle e_{\omega}, f_{FF}\rangle = 0. \label{lyap:2}
\end{align}
%
%
Thus, $H$ is a Lyapunov function about $(R_{d}, \omega_{d}^{s})$ and therefore $(R_{d},\omega_{d}^{s})$ is stable in the sense of Lyapunov for system (\ref{control_equation2}). $\blacksquare$\\

The next step is to introduce damping or the dissipative term $u_{diss}$ to achieve asymptotic stability. Introducing damping to the control by defining $u = (dV + f_{FF}) + u_{diss}$ 
where $u_{diss} = [u^{1}_{diss}~ u^{2}_{diss}~ u^{3}_{diss}]^{T}$. Then the closed loop control system becomes 
\begin{equation}\label{control_equation3}
\dot{q} = F_{cl}(q) + g(q)u_{diss}
\end{equation}
where $g = [g_{1}~g_{2}~g_{3}]$ and
\begin{align*}
F_{cl} & = \begin{bmatrix}
R_{s}\widehat{\omega}_{s}^{s} \\
M^{-1}\left( I^{s}\omega_{s}^{s} \times \omega_{s}^{s} + \mathbb{J}\dot{\Theta} \times \omega_{s}^{s} + (dV + f_{FF})\right) 
\end{bmatrix},~ g_{i} = \begin{bmatrix}
0 \\
-b_{i}
\end{bmatrix}.
\end{align*}
%
%
\textit{Lemma 1:} The control system (\ref{control_equation3}) is locally controllable on $SO(3) \times \mathbb{R}^{3}$.\\
\textit{Proof}: The proof is given in Appendix. $\blacksquare$ \\
We now prove asymptotic stability of our system about the desired equilibrium. To prove this we use the stability result stated in [\citep{bullo_stability},theorem 1].
\begin{thm}\label{theorem2}
Consider the system (\ref{control_equation3}) with input torque $u_{diss}$. Let $H$ be described in (\ref{lyapunov_function}). If $\mathcal{L}_{F_{cl}}H = 0$ and $u_{diss} = - \mathcal{L}_{g}H$ is the dissipative input, then the closed loop system asymptotically stabilize $(R_{d}, \omega_{d}^{s})$.
\end{thm}
\textit{Proof:} Consider the Lyapunov function $H$ as defined in (\ref{lyapunov_function}), Computing the rate of $H$ we get
\begin{align}\nonumber
\begin{split}
\frac{dH}{dt}& = \frac{\partial H}{\partial q} \dot{q} = \mathcal{L}_{F_{cl}}H + \mathcal{L}_{g}H u_{diss},
\end{split}
\end{align}
From (\ref{lyap:2}), we see that, $\mathcal{L}_{F_{cl}}H = 0$ which implies
\begin{equation}\label{asymptotic_H}
\dot{H} = \mathcal{L}_{g}H u_{diss}.
\end{equation}
%
Defining $u_{diss} = - \mathcal{L}_{g}H$ yield $\dot{H} = - (\mathcal{L}_{g}H)^{2}$. We know that $u_{diss}=\begin{bmatrix}
u_{diss}^{1} & u_{diss}^{2} & u_{diss}^{2}
\end{bmatrix}^{T}$ then calculating $u_{diss}^{i}$ as
%
\begin{align*}
u_{diss}^{i} = - \left(\frac{\partial H}{\partial q}\right)^{T}g_{i} = -\begin{bmatrix}
\left( \frac{\partial H}{\partial R_{s}}\right)^{T} & \left( \frac{\partial H}{\partial \omega_{s}^{s}}\right)^{T}
\end{bmatrix}
\begin{bmatrix}
\textbf{0} \\ 
-b_{i}
\end{bmatrix} = M e_{\omega} \cdot b_{i}
\end{align*}
where $i = {1,2,3}$. From this the dissipative control is calculated as
\begin{align*}
u_{diss} & = - \begin{bmatrix}
\mathcal{L}_{g_{1}}H & \mathcal{L}_{g_{2}}H & \mathcal{L}_{g_{3}}H
\end{bmatrix}^{T} = M e_{\omega} \cdot (M^{-1})^{T}= K_{v} e_{\omega},
\end{align*}
where $M=M^{T}$ is a symmetric positive-definite matrix and $K_{v}$ is a positive constant. Substituting the value of $\mathcal{L}_{g}H$ in (\ref{asymptotic_H}) we get $\dot{H} = - K_{v} (e_{\omega})^{2} \leq 0.$
Since, system is locally controllable from Lemma 1 and $\dot{H}$ is negative semidefinite we conclude from the Theorem 1 of \citep{bullo_stability} that the point $(R_{d}, \omega_{d}^{s})$ is local asymptotically stable. $\blacksquare$\\

%
\begin{corollary}
 If $\omega_{d}^{s} = 0$ implies $f_{FF} =0$, then the system (\ref{control_equation2}) with control $u = dV + K_{v}\omega_{s}^{s}$ is local asymptotically stable about $(R_{d},0)$.
\end{corollary}
Infact the system is local exponential stable about
$(R_{d},0)$. To check exponential stability, we compute the second variation of $H$. If the second variation is positive definite about the equilibrium point we say that the equilibrium is exponentially stable. From (\ref{lyapunov_function}) $H = T + V$ where $T= (1/2)\omega_{s}^{s}\cdot M \omega_{s}^{s}$ and for $T$ being a kinetic energy, yields $\partial^{2}T(q) > 0~ \forall q$. The second variation of the error functions $V$ is calculated as follows; let $\widehat{\eta} = R_{s}^{T}\delta R_{s} \in \mathfrak{so}(3)$, then
\begin{align}
\delta V(R_{s}) & = \delta (\mbox{trace}( K_{p}( I_{3\times 3} -  R_{s}R_{d}^{T})))=\mbox{trace}(\widehat{\eta} K_{p}  R_{s}R_{d}^{T}), \nonumber \\
\delta^{2} V(R_{s}) &= \delta \mbox{trace}(\widehat{\eta}  K_{p} R_{s}R_{d}^{T})= \mbox{trace}(- \widehat{\eta} K_{p}   \delta R_{s}R_{d}^{T}) \nonumber \\
&  = \langle \widehat{\eta},  K_{p}R_{s}R_{d}^{T} \widehat{\eta} \rangle \quad \quad \forall \widehat{\eta}\neq 0. \label{control_equation5}
\end{align}
At $R_{d}$ we have $\partial^{2}V>0$ and $\partial^{2}H(q_{0})= \partial^{2}T(q_{0}) + \partial^{2}V(q_{0})  > 0$, where $q_{0} = R_{d}$. Since, the second variation of $H$ is positive definite at equilibrium one can conclude the system achieves the desired orientation exponentially.

\subsubsection*{Simulation:}
\label{section_simu}
\begin{figure}
			\subfloat[]{
\includegraphics[scale=0.4]{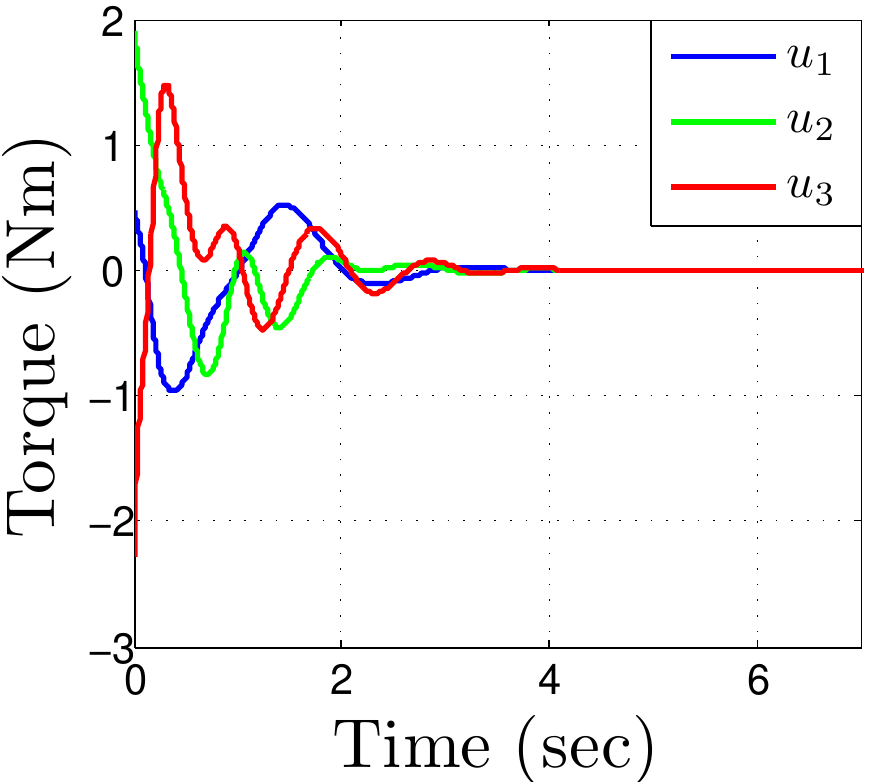}}			
			\subfloat[]{
\includegraphics[scale=0.38]{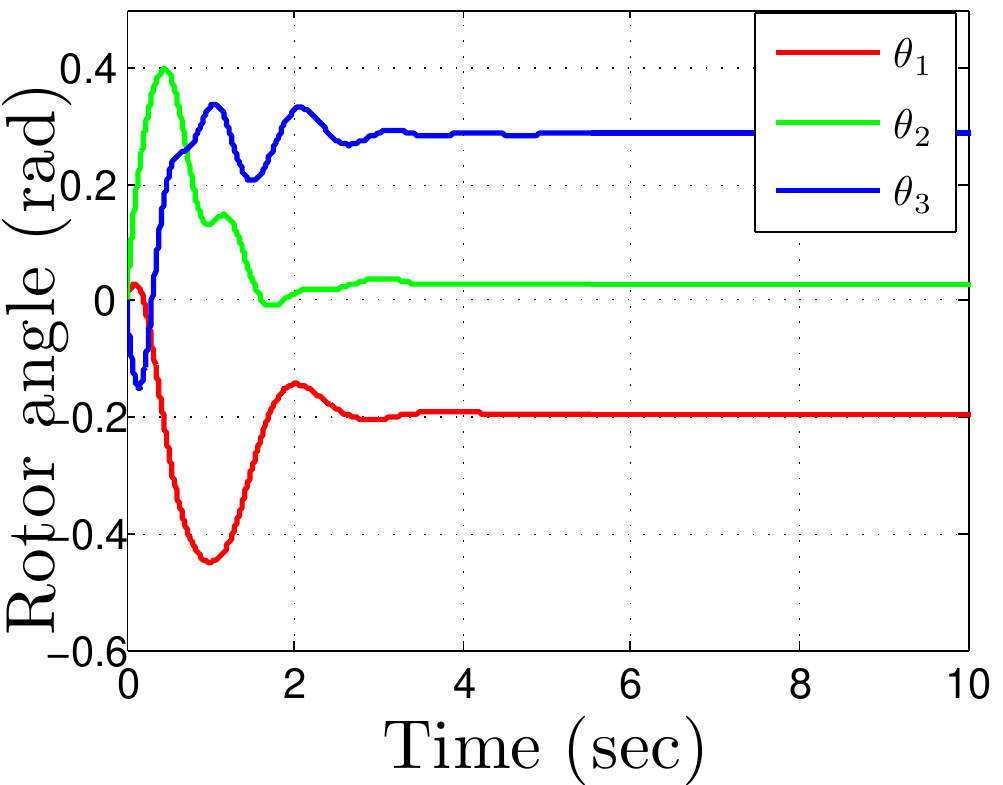}}
		\caption{(a) Torque to the internal rotors. (b) Angular position of the rotors. }				
				\label{control}				
\end{figure}

We choose the model parameters as: $m_{s} = 1kg;~ m_{i}= 0.672kg;~ r_s = 0.176m;~I^{s} = \hbox{diag}(0.0153,0.0153,$ $0.0153)kg-m^{2}$; $\mathbb{J}= \hbox{diag}(0.672,0.672,0.672)kg-cm^{2}$;
and control parameters as $K_{p} = \mbox{diag}(2,8,1)$ and $K_{v} = 0.5$. Choosing rotation matrix $R_{s}=\mbox{exp}(\alpha \hat{e}_{1})\mbox{exp}(\beta \hat{e}_{3})\mbox{exp}(\gamma \hat{e}_{1})$, the simulations are carried out for both attitude tracking and stabilization by three rotors with the following control law:
$$u = -(\sum_{i=1}^{3} \lambda_{i}R_{s}^{T}\hat{e}_{i} \times R_{d}^{T}\hat{e}_{i}) + K_{v}e_{\omega} + f_{FF}.$$ 
Keeping the desired orientation trajectory as $R_{d}(t) = exp(2 \pi (1 - \cos \pi t)\hat{e}_{2})$, then Fig. (\ref{error_angular_velocity}(a)) and (\ref{error_angular_velocity}(b)) shows the error in angular velocity ($e_{\omega}$) of the sphere and the error norm of $R_{s}$, indicating asymptotic convergence to the desired trajectory. The error norm is calculated as $$E_{R} = (3 - \mbox{trace}(K_{p}R_{d}^{T}R_{s}))^{1/2}.$$ 
For stabilization, setting the desired orientation as $R_{d}=R_{x}(\pi/9)R_{y}(\pi/18)R_{z}(\pi/3)$ and initial angular velocity as $\omega_{s}^{s}(0) = (12.5,7,1)$. 
Then Fig. (\ref{control}(a)) shows the torque applied at the internal rotors. The angular velocity of the three rotors is shown in Fig (\ref{control}(b)) converges to the initial momentum.
\begin{figure}[h]
			\subfloat[]{
\includegraphics[scale=0.41]{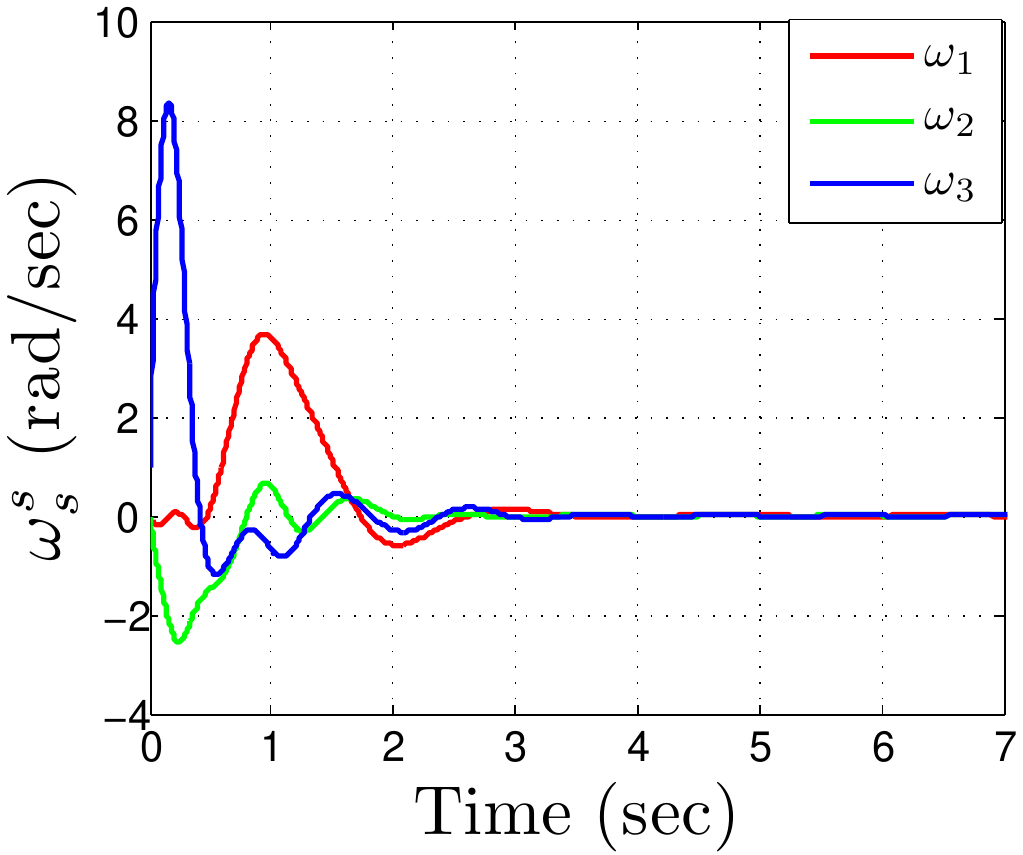}}			
			\subfloat[]{
\includegraphics[scale=0.43]{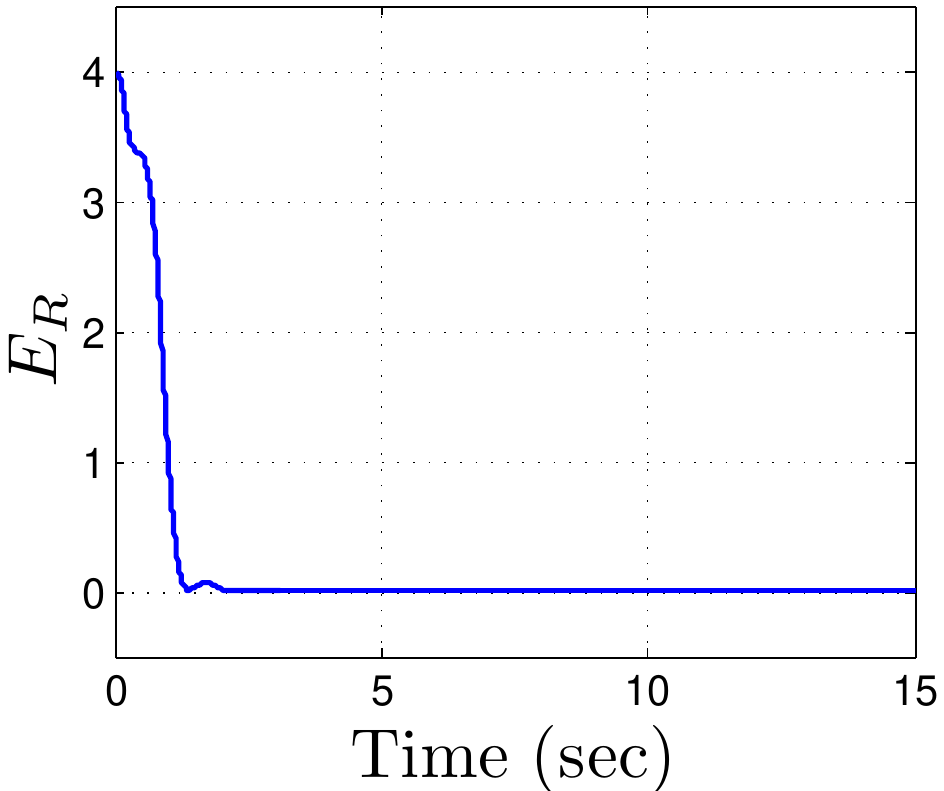}}
		\caption{(a)Angular velocity of spherical robot. (b) Error norm of orientation}				
				\label{angular_velocity}			
\end{figure}
%
%
%
%
%
%
%
\section{Contact point tracking and axis stabilization}
In this section we derive a control law based on a configuration error function which ensures contact position tracking by tracking the angular velocity. The control objective is to design a control law which aligns $\omega_{s}^{s}$ to a desired angular velocity and stabilizes/tracks the contact position on the plane asymptotically. Suppose $\dot{\mathbf{x}}_{d} = R_{d}\omega_{d}^{s} \times r e_{3}$ is the desired contact point velocity, where $R_{d}$ is a desired orientation. Note that given the nonholonomic constraint, $\dot{\mathbf{x}}_{d}$ gets determined by $\omega_{d}^{I} = R_{d} \omega_{d}^{s}$ which eventually determine $\mathbf{x}_{d}$ . Let $V_{1}$ be a potential function given by $V_{1} = \frac{1}{2}\| \mathbf{x} - \mathbf{x}_{d} \|^{2}.$
%
%
Taking the time derivative of $V$ along the system's trajectory,
\begin{align}\label{track_eq1}
\begin{split}
\dot{V}_{1} & = (\mathbf{x} - \mathbf{x}_{d}) \cdot (\dot{\mathbf{x}} - \dot{\mathbf{x}}_{d}) \\
& = (\mathbf{x} - \mathbf{x}_{d}) \cdot [(R_{s}\omega_{s}^{s} \times r \hat{e}_{3}) -  (R_{d}\omega_{d}^{s} \times r \hat{e}_{3})],\\
& = - r \hat{e}_{3} \times (\mathbf{x} - \mathbf{x}_{d}) \cdot R_{s}(\omega_{s}^{s} - R_{s}^{T}R_{d}\omega_{d}^{s}), \\
& =  - r R_{s}^{T}[\hat{e}_{3} \times (\mathbf{x} - \mathbf{x}_{d})] \cdot (\omega_{s}^{s} - R_{s}^{T}R_{d}\omega_{d}^{s}),
\end{split}
\end{align}
Set $R_{e} = R_{d}^{T}R_{s}$ and define the error in angular velocity as $e_{\omega} \triangleq \omega_{s}^{s} - R_{e}^{T}\omega_{d}^{s}$. The proportional-derivative (PD) control term is given as $f_{PD} = k_{p}r R_{s}^{T}[\hat{e}_{3} \times (\mathbf{x} - \mathbf{x}_{d})] - k_{d}e_{\omega}$, 
%
where $k_{p}$ and $k_{d}$ are positive definite matrices. With this PD control and (\ref{feedfrwd}), the nonlinear controller is given as
\begin{equation}\label{control_track}
u = -(f_{PD} + f_{FF}).
\end{equation}
\begin{thm}
The system (\ref{reduced_dyanamic}) with control input (\ref{control_track}), given by
\begin{equation}\nonumber
\dot{\omega}_{s}^{s} = M^{-1}(I^{s} \omega_{s}^{s} + \mathbb{J}\dot{\Theta}) \times \omega_{s}^{s} + M^{-1}(f_{PD} +  f_{FF}) ,
\end{equation}
is local asymptotically stable at $(\mathbf{x}_{d},\omega_{d}^{s})$.
\end{thm}
\textit{Proof}: Define a candidate error function
\begin{equation}
H(\mathbf{x},\omega_{s}^{s}) = V_{1} + \frac{1}{2}\| e_{\omega}\|_{M}^{2} = V_{1} + \frac{1}{2}\mathbb{G}(I)(e_{\omega},e_{\omega}),
\end{equation}
The time derivative of the Lyapunov function is
\begin{align}
& \frac{d}{dt}H(\mathbf{x},\omega_{s}^{s}) = \frac{d}{dt}V_{1} + \mathbb{G}(I)(e_{\omega},\nabla_{\omega_{s}^{s}}e_{\omega}), \nonumber \\
& = \dot{V}_{1} + \mathbb{G}(I)(e_{\omega},\nabla_{\omega_{s}^{s}}\omega_{s}^{s} - \nabla_{\omega_{s}^{s}}R_{e}^{T} \omega_{d}^{s}), \nonumber \\
& = - r R_{s}^{T} [\hat{e}_{3} \times (\mathbf{x} - \mathbf{x}_{d})]\cdot e_{\omega} + \langle e_{\omega}, f_{PD} + f_{FF}  \rangle - \langle e_{\omega}, f_{FF}\rangle, \nonumber \\
& = - k_{d} e_{\omega}\cdot e_{\omega} \leq 0. \label{lyap:1}
\end{align}
Let $\Omega_{c} = \{ (\mathbf{x}, \omega_{s}^{s}) | H(\mathbf{x},\omega_{s}^{s}) \leq c \}$,where $c>0$ and since $\dot{H}(\mathbf{x},\omega_{s}^{s}) \leq 0$ all the trajectories are bounded and contained within $\Omega_{c}$. Define $N$ to be the set of all points  of $\Omega_{c}$ satisfying $\dot{H}=0$. From (\ref{lyap:1}), we have $N = \{(\mathbf{x},\omega_{s}^{s})\in \Omega_{c} | e_{\omega}=0 \}$. As $e_{\omega}=0$ implies $\omega_{s}^{s} = R_{e}^{T}\omega_{d}^{s}$ which yields the dynamics, $\dot{\mathbf{x}} = \dot{\mathbf{x}}_{d}$ and $\dot{\omega}_{s}^{s} = - r M^{-1}R_{s}^{T}[\hat{e}_{3} \times (\mathbf{x} - \mathbf{x}_{d})] + \frac{d}{dt}(R_{e}^{T}\omega_{d}^{s})$. Since the robot rolls on a horizontal plane, at any point $(\mathbf{x} - \mathbf{x}_{d}) \neq \hat{e}_{3}$. So the only possibility of $\dot{e}_{\omega} = 0$ to happen is when $x = x_{d}$. Hence, the largest invariant set will be the set $N_{1} = \{(\mathbf{x}, e_{\omega})| \omega_{s}^{s}=R_{e}^{T}\omega_{d}^{s}, \mathbf{x}=\mathbf{x}_{d}\}$ in $\Omega_{c}$. And from LaSalle's invariance principle, the trajectories in $\Omega_{c}$ converge to $N_{1}$ as $t \rightarrow \infty $, i.e, to the equilibrium $(\mathbf{x}_{d},\omega_{d}^{s})$. $\blacksquare$
\subsubsection*{Position and reduced attitude stabilization:} 
For position stabilization there are two cases: 1) when $x_{d}=0$ $\& $ $R_{d}\omega_{d}^{s} = 0$ $\Rightarrow$ $\omega_{d}^{s}=0$; and 2) when $x_{d}=0$ $\& $ $R_{d}\omega_{d}^{s} = \alpha \hat{e}_{3}$, where $\alpha$ is any scalar. Case 1 is immediate. When $\omega_{d}^{s}=0$ the control law $u = f_{PD}$ and the robot converges to the origin asymptotically. In case 2, $R_{d}\omega_{d}^{s} = \alpha \hat{e}_{3}$ implies that the robot's final contact position is the origin and the angular velocity is about the $z-$axis. The control law in this case is expressed as 
\begin{align}\nonumber
\begin{split}
u & = - k_{p}r R_{s}^{T}(\hat{e}_{3} \times \mathbf{x}) + k_{d} (\omega_{s}^{s} - \alpha R_{s}^{T}\hat{e}_{3}) - \frac{1}{2} (\omega_{s}^{s} \times \alpha M R_{s}^{T}\hat{e}_{3}) \\
& + \frac{1}{2} (M\omega_{s} \times \alpha R_{s}^{T}\hat{e}_{3}) - M (\omega_{s}^{s} \times \alpha R_{s}^{T}\hat{e}_{3}).
\end{split}
\end{align}
The first term in the control law is responsible for the contact position stabilization and remaining terms will orient the sphere such that the angular velocity tracks $\hat{e}_{3}$. Such is a case of reduced attitude stabilization where stabilizing $R_{s}$ upto a rotation about $\hat{e}_{3}$ is equivalent to stabilizing the angular velocity direction of the axis $R_{s}^{T}\hat{e}_{3}$ \citep{bullo_murray_track}. Thus, we can restate the attitude as $R_{s} \in \mathbb{S}^{2}$ and conclude that the control law (\ref{control_track}) gives the contact point and reduced attitude stabilization in terms of the points in $\mathbb{R}^{2} \times \mathbb{S}^{2}$. 
\subsubsection*{Simulation:}
\begin{figure}[h]
\centering
			\subfloat[]{
\includegraphics[scale=0.39]{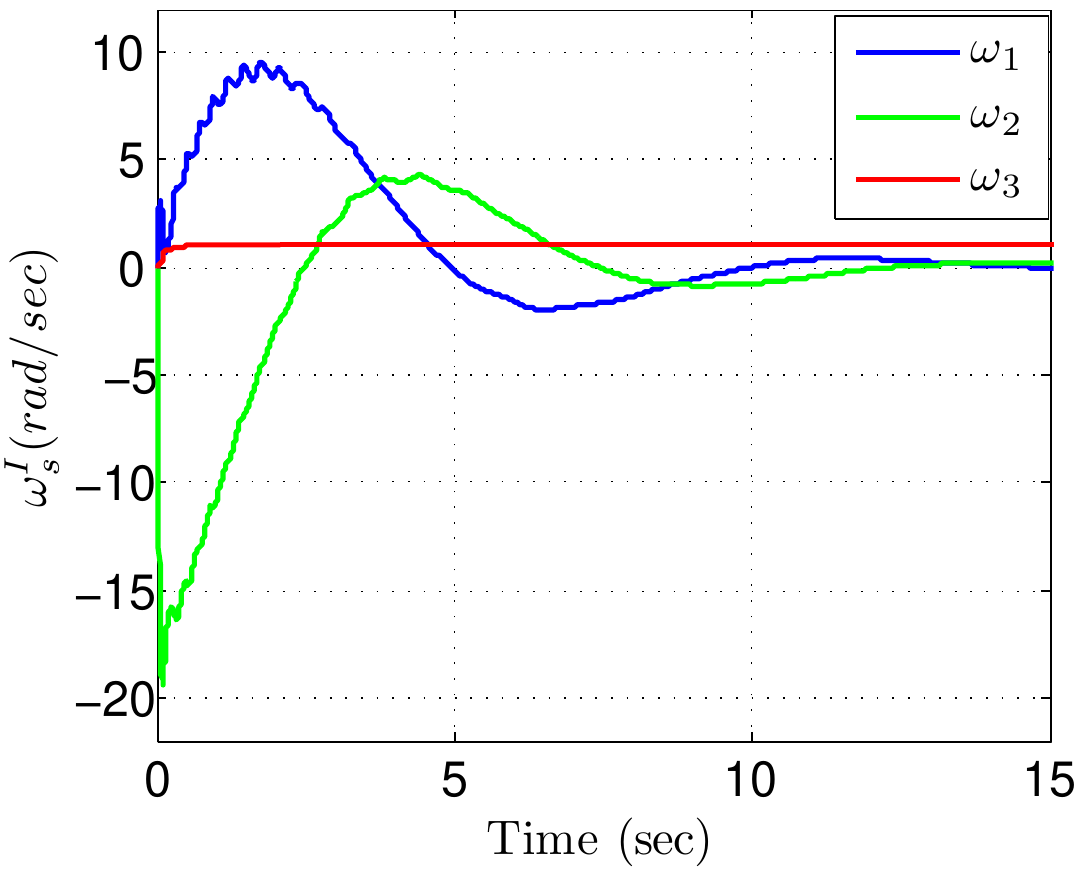}}			
			\subfloat[]{
\includegraphics[scale=0.45]{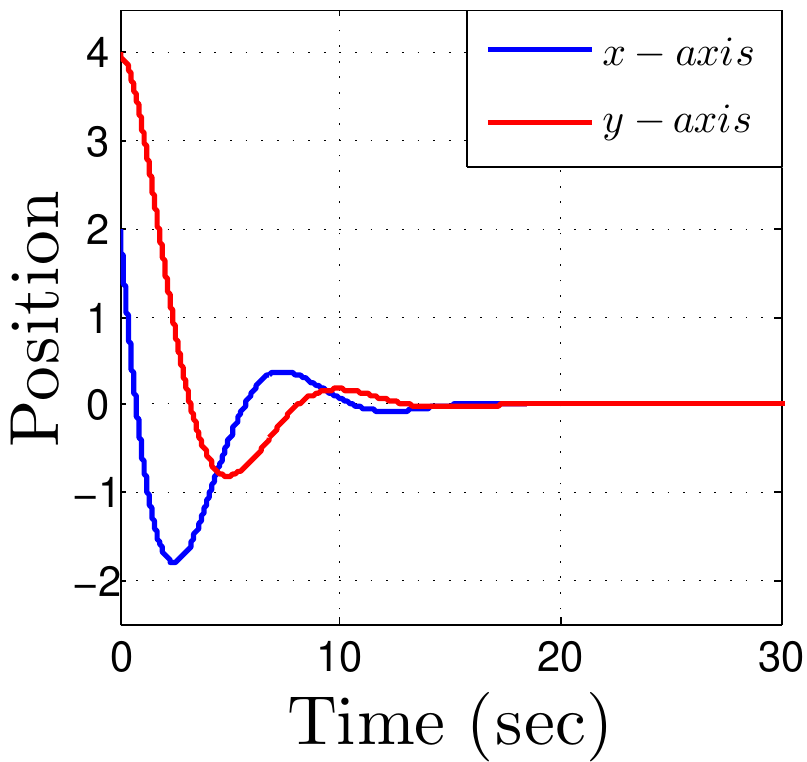}}
		\caption{a)Angular velocity of spherical robot. b) Position on the plane.}				
		\label{position_track1}				
\end{figure}

\begin{figure}[h]
\centering
			\subfloat[]{
\includegraphics[scale=0.45]{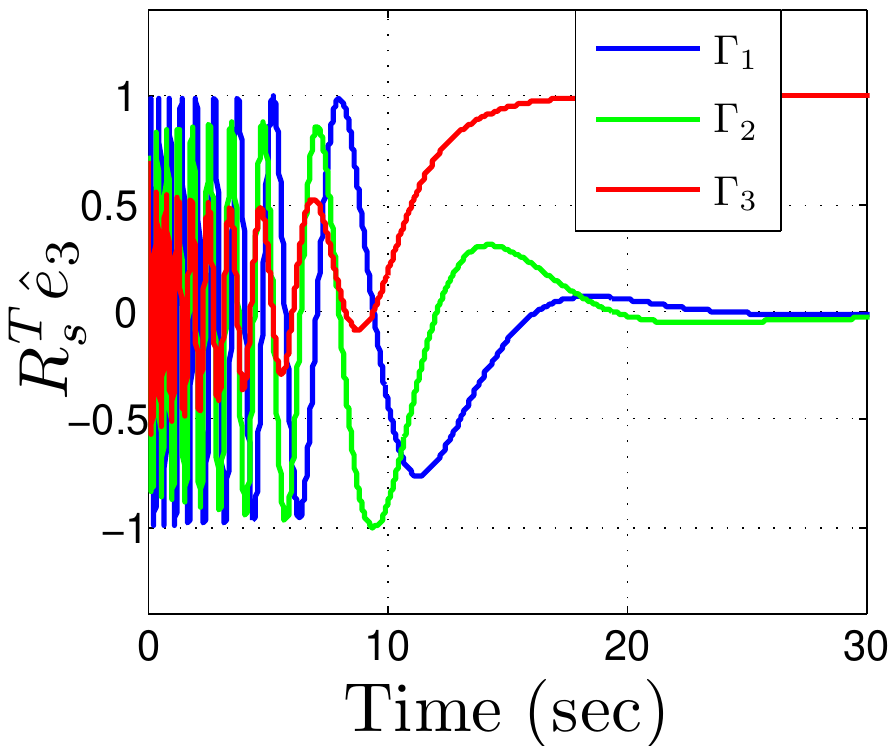}}			
			\subfloat[]{
\includegraphics[scale=0.45]{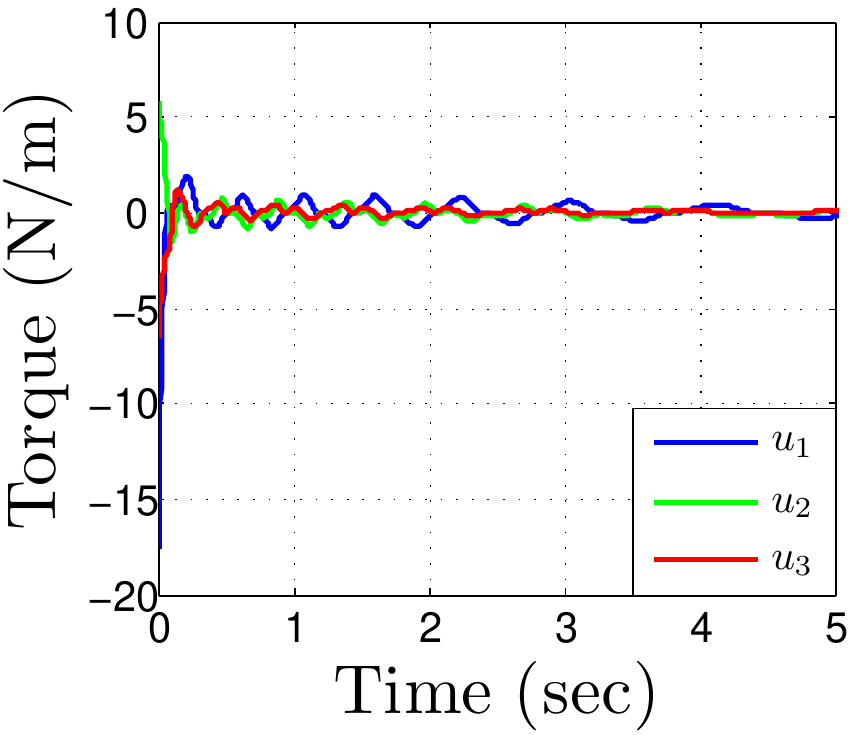}}
			\caption{ $\Gamma$ plot and torques on rotors.}				
				\label{fig:gam}
\end{figure}

\begin{figure}[h]
\centering
			\subfloat[]{
\includegraphics[scale=0.47]{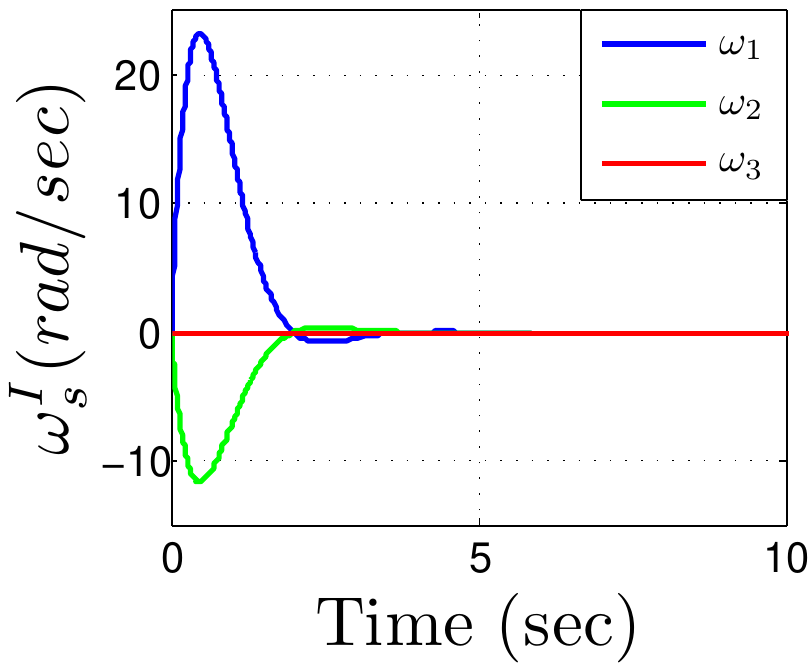}}			
			\subfloat[]{
\includegraphics[scale=0.45]{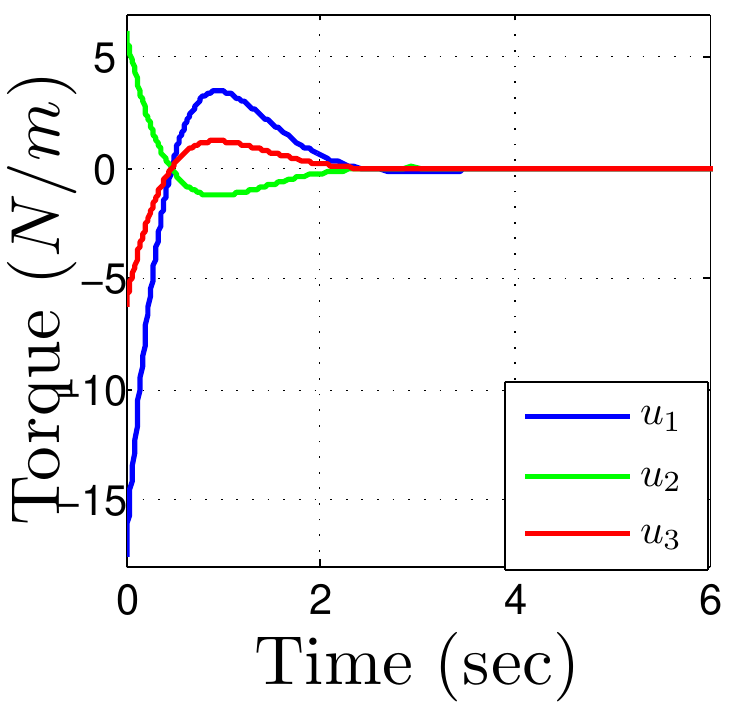}}
\caption{Angular velocity and torques on rotors.}				
\label{fig:5}
\end{figure}

\begin{figure}[h]
\centering
			\subfloat[]{
\includegraphics[scale=0.4]{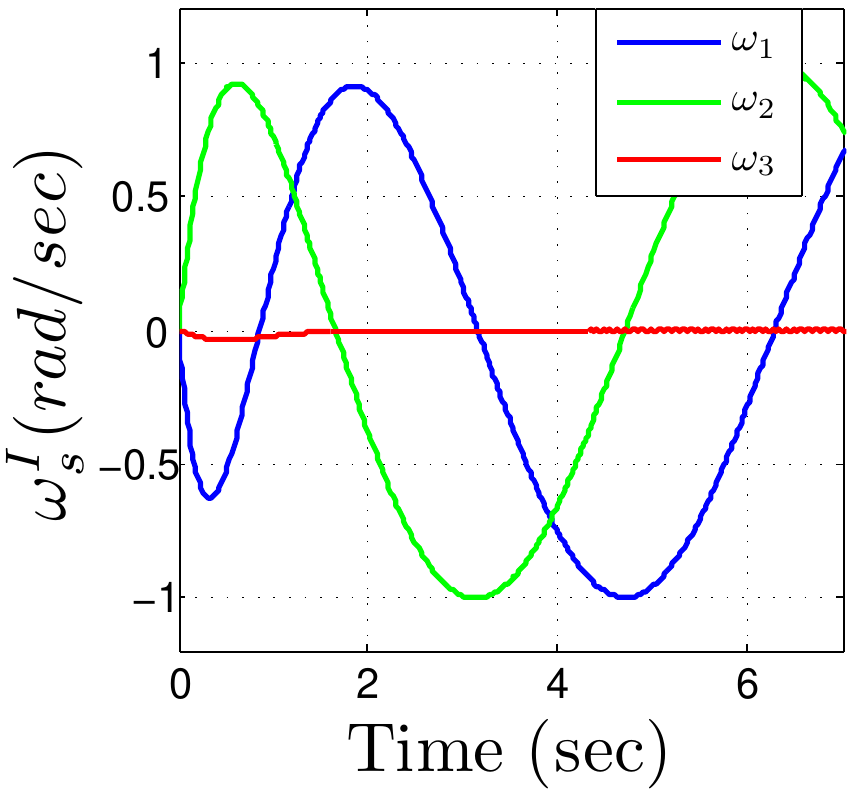}}			
			\subfloat[]{
\includegraphics[scale=0.5]{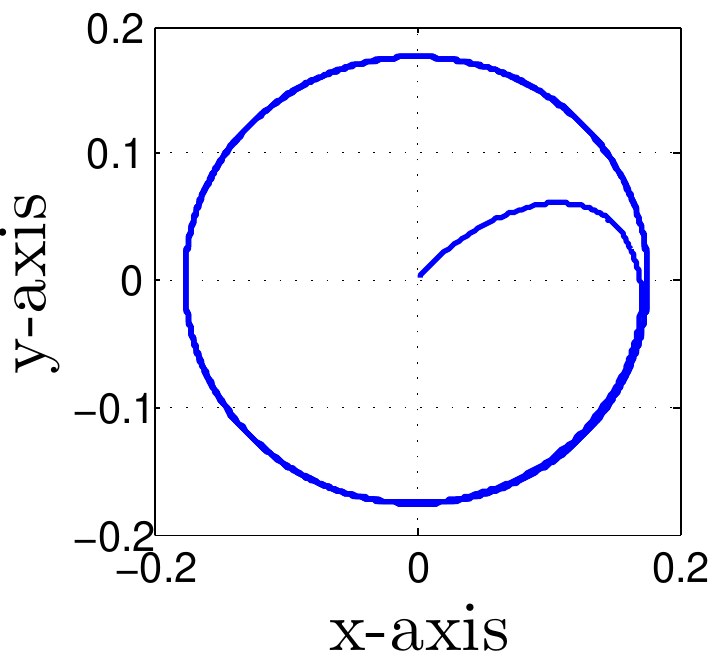}}
\caption{Angular velocity and phase plane of $xy$ position tracks the circular trajectory.}			
\label{fig:6}
\end{figure}

\begin{figure}[h]
\centering
			\subfloat[]{
\includegraphics[scale=0.48]{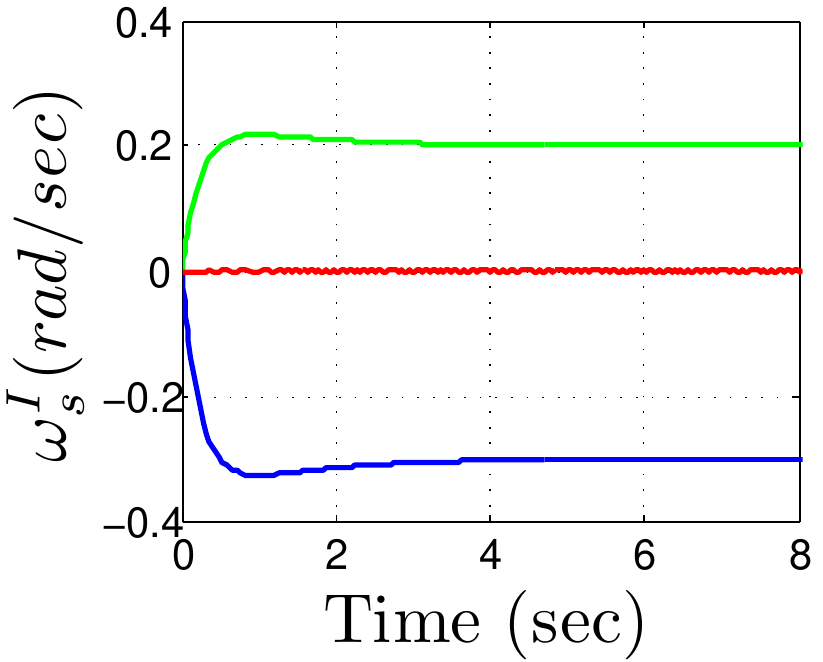}}			
			\subfloat[]{
\includegraphics[scale=0.5]{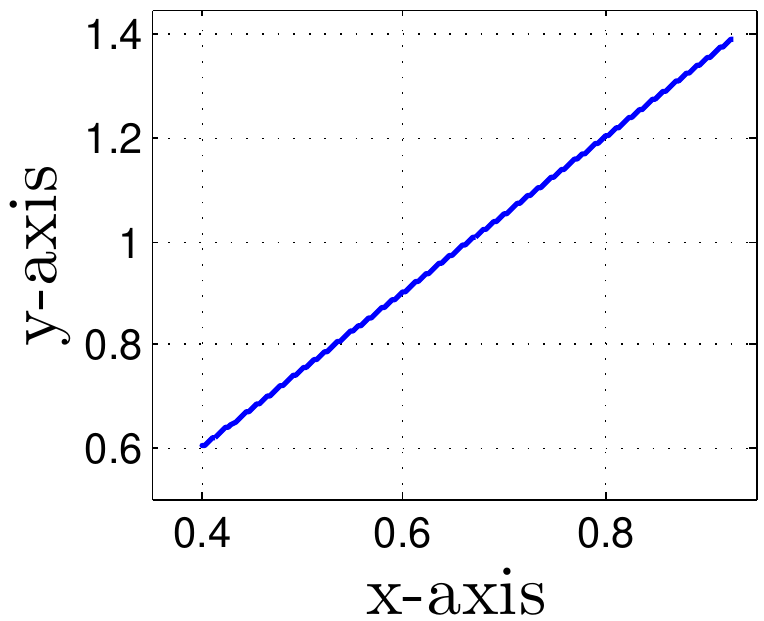}}
\caption{Angular velocity and phase plane of $xy$ position tracks the line trajectory.}			
\label{fig:7}
\end{figure}

We take the model parameters as in section (\ref{section_simu}) with initial orientation $R_{s_{0}} = exp(\frac{\pi }{6} \hat{e}_{1}) $ and starting point on the horizontal plane as $(x_{0},y_{0}) = (4,2)\mbox{units}$. Setting the desired orientation $R_{d} = exp(\frac{\pi}{4} \hat{e}_{3})$ and $\omega_{d}^{s} = \hat{e}_{3}$ which satisfy $R_{d}^{T}\omega_{d}^{s}=\hat{e}_{3}$, Fig. \ref{position_track1}(a)) and Fig. (\ref{fig:gam}(a)) shows that as the angular velocity achieve $\omega_{d}^{s}$ asymptotically, the sphere attains the desired Line-of-sight that is $R_{s}^{T}\hat{e}_{3}= (\Gamma_{1},\Gamma_{2},\Gamma_{3})$ converges to $\hat{e}_{3}=(0,0,1)$. The initial oscillations in $R_{s}^{T}\hat{e}_{3}$ plot are due to the sphere rotating in the spiral type motion on plane and then asymptotic converges to $(0,0,1)$. The position on $xy$ plane is illustrated in Fig. (\ref{position_track1}(b)).
To illustrate contact point trajectory tracking, we choose $\mathbf{x}_{d}$ to track line and circle, as shown in Fig. (\ref{fig:5}), (\ref{fig:6}) and (\ref{fig:7}). Keeping $\omega_{d}^{s}=0$, then Fig. (\ref{fig:5})(a) and (b) shows the angular velocity of the sphere and the phase plane of position. Setting $\mathbf{x}_{d} = (r_{s}\sin (t),r_s \cos (t))$ which yields $\omega_{d}^{s}=(-\sin(t),-\cos(t))$. Fig. (\ref{fig:6}) shows the spherical robot follows the desired circular trajectory. Setting $\omega_{d}^{s}= (0.2,0.3)$ and $\mathbf{x}_{d} = (0.2t + 0.4,0.3t+0.6)$ then the sphere will rotate at constant speed shown in Fig. (\ref{fig:7}) and follows the line given by $\mathbf{x}_{d}$.
\begin{figure}[h]
\centering
\includegraphics[scale=0.35]{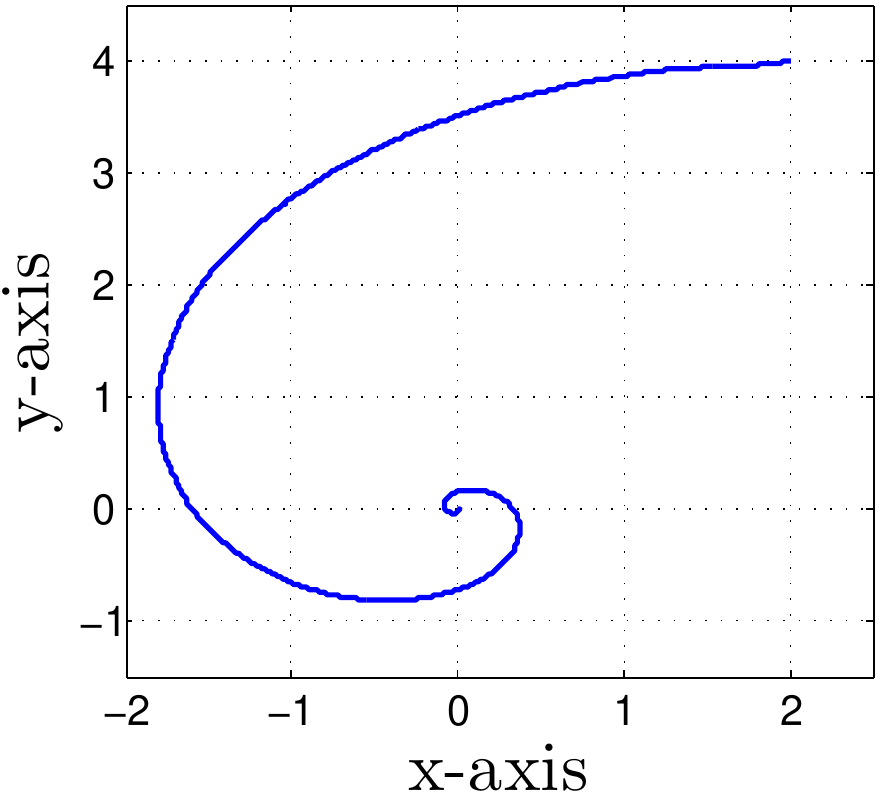}
\caption{Phase plane of $xy$ position.}			
\label{fig:4}
\end{figure}
\section{Discussion}
In conclusion, we say that both the control strategies derived using the geometric approach, without parametrization,  illustrate a more general philosophy on the control design, preserving the mechanical notions of the system. To the best of our knowledge this is the first instance where such a strategy has been employed to a nonholonomic system. Both the control  strategies are derived using the notions of the affine connection, the error functions and a transport map on tangent spaces. The first feedback strategy results in a continuous feedback law which tracks the desired orientation trajectory. In position tracking strategy, an intermediate result while proving the stability is $\nabla_{\omega_{s}^{s}}e_{\omega} = f_{PD}$, which provides an interpretation about feedforward control $f_{FF}$. The closed-loop system with $f_{FF}$ has the property that $\nabla_{\omega_{s}^{s}}e_{\omega}$ vanishes along the trajectory. That is, if $\e_{\omega} =0 \Rightarrow \dot{e}_{x}=(\dot{\mathbf{x}} - \dot{\mathbf{x}}_{d})$ is zero at initial time, it will remain zero at final time. And if we keep $\omega_{d}^{s}=\hat{e}_{3}$ we get the $\dot{e}_{x}(0)=\dot{e}_{x}(T)=0$ for all time, and the result is contact position and a reduced attitude stabilization. 
\section*{Appendix}
\subsubsection{Proof of Lemma 1:} 
\label{A3}
In this section we will compute the Lie brackets of $F_{cl}$ and $g_{i}$, where $i=1,2,3$. Given two vector field $X,Y \in TQ$ the Lie derivative (bracket) of $Y$ along $X$ is $[X,Y] \equiv \frac{d}{dt}|_{t=0}\Phi_{t}^{*}(Y)$, 
where $\Phi$ is the flow of $X$ and $\Phi_{t}^{*}(Y)$ is pull-back of a vector field $Y$. From system (\ref{control_equation3}), the Lie bracket of $F_{cl}$ and $g_{i}$ will be
\begin{equation}\nonumber
[F_{cl},g_{i}](q) = -[g_{i},F_{cl}](q) = - \frac{d}{dt}|_{t=0} (D\Phi_{t}^{g_{i}}(q))^{-1}\cdot F_{cl}(\Phi_{t}^{g_{i}}(q))
\end{equation}
where $\Phi_{t}^{g_{i}}$ is the flow of $g_{i}$. The control vector field $g_{i}(q) = (\mathbf{0}, M^{-1}\hat{e}_{i})$  
then flow of $g_{i}$ is given as $\Phi_{t}^{g_{i}}(q) = \left(R_{s},\omega_{s}^{s} + t M^{-1}\hat{e}_{i}  \right)$ 
%
and $(D\Phi_{t}^{g_{i}}(q))^{-1}$ is the identity map on the manifold $Q$.
\begin{align*}
& [g_{1},F_{cl}](q) = \frac{d}{dt}|_{t=0} (D\Phi_{t}^{g_{i}}(q))^{-1}\cdot F_{cl}(\Phi_{t}^{g_{i}}(q)) \nonumber \\
& = \frac{d}{dt}|_{t=0} (D\Phi_{t}^{g_{i}}(q))^{-1}\cdot \begin{bmatrix}
R_{s}(\omega_{s} + t M^{-1}\hat{e}_{1})^{\bigwedge} \\
\ast
\end{bmatrix} = \begin{bmatrix}
R_{s}(M^{-1}\hat{e}_{1})^{\wedge}\\
\ast
\end{bmatrix}.
\end{align*}
Similarly, $[g_{2},F_{cl}](q)$ and $[g_{3},F_{cl}](q)$ are calculated and given as
\begin{equation*}
[g_{2},F_{cl}] = \begin{bmatrix}
R_{s}(M^{-1}\hat{e}_{2})^{\wedge}\\
\ast
\end{bmatrix}\mbox{   and   }
[g_{3},F_{cl}] = \begin{bmatrix}
R_{s}(M^{-1}\hat{e}_{3})^{\wedge}\\
\ast
\end{bmatrix}.
\end{equation*}
where $\ast$ denotes some functions we are not interested in. The vectors $g_{1}$,$g_{2}$,$g_{3}\in T_{q}Q$ are linearly independent since $\{\hat{e}_{1},\hat{e}_{2},\hat{e}_{3}\}$ are linearly independent. To see the linear independence, we write all the six vectors as
\begin{align}
& \alpha_{i}M^{-1}\hat{e}_{i} + \beta_{i}R_{s}(M^{-1}\hat{e}_{i})^{\wedge}=0. \label{lin_indep2}
\end{align}
Now, for these vectors to be linearly independent, all the scalars  $\alpha_{i}$'s and $\beta_{i}$'s equal to zero. From the values of $g_{i}$ and  $[g_{i},F_{cl}]$, for (\ref{lin_indep2}) to hold  that, it follows that
\begin{align}\label{lin_indep1}
\alpha_{i}M^{-1}\hat{e}_{i}  = 0 \quad \beta_{i}R_{s}(M^{-1}\hat{e}_{i})^{\wedge}=0. 
\end{align}
for all $i$. Since $\{g_{1},g_{2},g_{3}\}$ is linearly independent, (\ref{lin_indep1}) will hold only when $\alpha_{i}=0$.  And $\{R_{s}(M^{-1}\hat{e}_{1})^{\wedge},R_{s}(M^{-1}\hat{e}_{2})^{\wedge},$  $R_{s}(M^{-1}\hat{e}_{3})^{\wedge}\}$ is linear independent, then $\beta_{i}=0$ to satisfy (\ref{lin_indep2}). Hence, the set
$\{ g_{1},g_{2},g_{3},[f,g_{1}],[f,g_{2}],[f,g_{3}]\}$ 
are linearly independent on $Q =SO(3)\times \mathbb{R}^{3}$ of dimensional six and spans the tangent space of the configuration space at any configuration. Therefore, the system is locally controllable. 
%

                                                   







\end{document}